\documentclass[preprintnumbers,amsmath,11pt,amssymb,floatfix,superscriptaddress,nofootinbib]{revtex4}
\usepackage{graphicx}
\usepackage{epsfig}
\usepackage{bm}
\usepackage{amsfonts}

\input epsf

\begin{document}

\title{\Large Reconstruction of $f(G)$ gravity with ordinary and entropy-corrected $(m,n)$ type Holographic dark energy model}

\author{Rahul Ghosh}
\email{ghoshrahul3@gmail.com} \affiliation{ Department of
Mathematics, Bhairab Ganguly College, Kolkata-700 056, India.}

\author{Ujjal Debnath}
\email{ujjaldebnath@gmail.com} \affiliation{Department of
Mathematics, Bengal Engineering and Science University, Shibpur,
Howrah-711103, India}

\date{\today}

\begin{abstract}
We have discussed the correspondence of the well-accepted $f(G)$
gravity theory with two dark energy models : $(m,n)$-type
holographic dark energy [$(m,n)$-type HDE] and entropy-corrected
$(m,n)$-type holographic dark energy. For this purpose, we have
considered the power law form of the scale factor
$a(t)=a_{0}t^{p}, p>1$. The reconstructed $f(G)$ in these models
have been found and the models in both cases are found to be
realistic. We have also discussed the classical stability issues
in both models. The $(m,n)$-type HDE and its entropy-corrected
versions are more stable than the ordinary HDE model.
\end{abstract}

\pacs{}

\maketitle

\section{Introduction} The recent observational data from Ia
supernovae, the large scale structure and cosmic microwave
background anisotropies confirm that the universe is undergoing a
late-time acceleration \cite{1,2,3,4,5,6,7,8,9,10,11}. A unknown
component, known as dark energy, is believed to be responsible for
this accelerated expansion. Cosmological observation suggest that
the two-thirds of the the total energy of the universe is been
occupied by this dark energy. Remaining portion is almost occupied
by dark matter with a little presence of baryonic matter
\cite{12}.\\
There are many candidates for dark energy and the simplest among
them is a tiny positive cosmological constant with equation of
state parameter $\omega=-1$. But the fine-tuning and the
cosmological coincidence problem occur when we deal with the
cosmological constant. Hence we tend to look at the dynamic scalar
field models whose equation of state parameter are not constant
but evolve with the cosmic time. The modified gravity approach is
a gravitational alternative to explain the accelerated expansion.
These two approaches may compliment each other as shown by the
works of many authors
\cite{13,14,15,16,17,18,19,20,21,22,23,24,25,26,27,28,29}.\\
Holographic principle proposed by Fischler and Suskind \cite{30}
is one of the important results of the recent investigations that
arise exploring the quantum gravity theory or the string theory
\cite{31,32}. Depending on this holographic principle the nature
of the dark energy in the context of quantum gravity is called the
holographic dark energy (HDE). An enlargement relationship as
proposed by \emph{Cohen et al} \cite{33} between the infra-red
(IR) and ultraviolet cut-offs due to the limitation set by the
formation of a black-hole, which establishes an upper limit for
the vacuum energy, $L^{3}\rho_{v}\leq L{M_{P}}^2 $, where
$\rho_{v}$ is the HDE density related to the UV cut-off, $L$ is
the IR cut-off and $M_{P}$ is the reduced Planck mass. The HDE may
provide simultaneously natural solutions to both dark energy and
cosmological problems \cite{34}. Specially it can resolve the
coincidence problem \cite{35} and the phantom crossing \cite{36}.
Besides the model has reasonable agreement with the astrophysical
data of CMB, SNeIa and galaxy redshift
   surveys \cite{37}. On these accounts, HDE paradigm has been extended via
   different cut-offs \cite{38} and entropy corrections \cite{39}. The holographic scale L can
 be identified with the future event horizon [34] ,
   the conformal age of the universe  \cite{34A} or the Ricci scalar of the universe  \cite{34B}.
   Specially the model taking the conformal age of the universe is also dubbed as new agegraphic dark
   energy model. A recent extension
   of this idea is $(m,n)$-type HDE \cite{40}, where $m$ and $n$ are the
   parameters associated with the chosen IR cut-off
   \begin{equation}
   L=\frac{1}{a^{m}(t)}\int_{0}^{t}
   a^{n}(t')dt'
   \end{equation}
   A holographic dark energy model with a conformal-like age of the universe as
   the scale L  \cite{34C} is consistent with the history from the inflation to the current
universe. In this sense, all such scales are proposed at a
phenomenological level. In analogy with the trend, perhaps the
direct physical motivation of proposing such characteristic scales
for $(m,n)$-type holographic dark energy model is still obscure,
but it generalized the theory with significant improvements and
the parameters $(m, n)$ provide us more space in theory to fit the
observational data. In particular, for some specific values of
$(m, n)$ the equation of state can naturally evolve cross phantom
divide even without introducing an interaction between dark energy
and dark matter. Also when $(m, n)$ take some specific value all
the agegraphic-like dark energy models can be recovered. This
construction is also applicable to the models with generalized
future event horizon as the holographic size in the same spirit.
For age-like holographic models, when $m = n$  it seems that dark
energy has the same behavior as the dominant ingredient in the
early epoches of the universe. Which  imply  that dark energy
might be unified with dark matter. However, we have to introduce
some mechanism to make dark energy deviate from dark matter state,
and eventually become dominant  and  be responsible for the
acceleration of the universe. To achieve this we need some
 appropriate interactions between dark energy and dark matter.
This model is always stable in the dark energy dominated era in
Newtonian gauge. The preliminary numerical analysis has indicated
that this model fits with the observational data very well. The
best-fit analysis in this reference indicates that this model with
$m = n+1$ and small m is more favored including the cases of
$(m,n)=(1,0),(2,1),(3,2),(4,3)$ etc \cite{34D}.\\
 Black-hole entropy (BH) $S_{BH}$ plays an important role in
 the derivation of HDE. It is well-known that
   $S_{BH}=\frac{A}{4G}$, where $A(\sim L^{2})$ is the area of
   the BH horizon. The BH entropy-area relation faces a
   modification in loop quantum gravity (LQG) due to thermal
   equilibrium fluctuation, quantum fluctuation or mass and charge
   fluctuations, in the form \cite{44,45,46} $S_{BH}=\frac{A}{4G}+\xi\log
   \frac{A}{4G}+\zeta$, where $\xi$ and $\zeta$ are dimensionless
   constants of unit order. Wei \cite{47} has proposed the energy-density of
   the entropy-corrected HDE with the help of this corrected
   entropy-area relation setup of LQG and obtained the energy
   density of the entropy-corrected HDE. The application of the
   the IR cut-off $L=\frac{1}{a^{m}(t)}\int_{0}^{t}
   a^{n}(t')dt'$ here would be a interesting aspect to see.
   \\Modification of the Einstein's general theory of relativity
   is popular aspect to explain the the accelerated cosmic
   expansion. $f(R)$ gravity \cite{48,49,50A}, $f(T)$ gravity
   \cite{50B,50C}, $f(G)$ gravity \cite{51}, $f(R,T)$
   gravity, $f(R,G)$ gravity are some popular modified gravity
   models. Beside explaining cosmic expansion these theories are
   also helpful in explaining the unification of early-time
   inflation and late-time acceleration era and then the
   non-phantom to phantom phase and many more. There are many
   works in the literature on the correspondence between different
   kinds of HDE and modified gravity \cite{26,27,52,53,54,55,56}. In this work, we have
   considered the correspondence of $f(G)$ gravity (where $G$
   is the Gauss-Bonnet invariant) with ordinary and entropy-corrected $(m,n)$-type HDE in
   the light of power law of $a(t)$.\\
The Einstein's field equations in FRW model in the background of
$f(G)$ gravity are discussed in section II. Our main aim is to
discuss the correspondence of the well-accepted $f(G)$ gravity
theory with two dark energy models : $(m,n)$-type holographic dark
energy [$(m,n)$-type HDE] and entropy-corrected $(m,n)$-type
holographic dark energy, which will be discussed in sections IIA
and IIB. We have also discussed the classical stability issues in
both models in section III. Finally, we draw some concluding
remarks.\\

\section{ \textbf{ Reconstruction of $f(G)$ gravity }}

The action of $f(G)$ is considered as \cite{57}
\begin{equation}
S=\int
d^{4}x\sqrt{-g}\left(\frac{1}{2}R+f(G)+\mathcal{L}_{m}\right)
\end{equation}
where, $f(G)$ is an arbitrary function of Gauss-Bonnet invariant
$G$,
$G=R_{\alpha\beta\gamma\delta}R^{\alpha\beta\gamma\delta}-4R_{\alpha\beta}R^{\alpha\beta}+R^{2}$,
and $\mathcal{L}_{m}$ is Lagrangian of the matter present in the
universe. This action describes the Einstein's gravity coupled
with perfect fluid plus a Gauss-Bonnet $(G)$ term. We set the
Planck mass to unity. The field equations in FLRW space-time with
signature $(-,+,+,+)$ are
\begin{equation}
3H^{2}=-24H^{3}\dot{f_{G}}+Gf_{G}-f+\rho_{m}=\rho_{eff}
\end{equation}
\begin{equation}
-2\dot{H}=-8H^{3}\dot{f_{G}}+16H\dot{H}\dot{f_{G}}+8H^{2}\ddot{f_{G}}+\rho_{m}=\rho_{eff}+p_{eff}
\end{equation}
 ~~~~where $-24H^{3}\dot{f_{G}}+Gf_{G}-f$ is the energy density
 contribution of $f(G)$, further $\rho_{eff}$ and $p_{eff}$ are
 the effective energy density and pressure respectively. Here
 $G=24H^2(\dot{H}+H^{2})$, $H=\frac{\dot{a}}{a}$, where the dot
 represents the time derivative. Now, further calculation reveals
 that $\rho_{m}=\rho_{0}a^{-3(1+\omega)}$, where $\rho_{0}$ is
 the matter density at redshift $z=0$ and $\omega$ is EoS
 parameter of the matter field.
 \\~~~~~~In literature, there is no unique functional form of
 $f(G)$, but to study the evolution, a functional form of $f(G)$
 is required. Now to reconstruct $f(G)$ gravity model we are
 trying to consider two models (i) $(m,n)$-type HDE Model and
 (ii) entropy-corrected $(m,n)$-type HDE Model.

\subsection{\textbf{Reconstruction with $(m,n)$-type HDE Model}}

Now to process to reconstruct $f(G)$ gravity with $(m,n)$-type
HDE, we need the energy density, which is given by \cite{40}
\begin{equation}
\rho_{v}=\frac{3b^{2}}{L^{2}}
\end{equation}
with $b=$ constant and $L=$ generalized IR cut-off defined as
\begin{equation}
L=\frac{1}{a^{m}(t)}\int_{0}^{t}
   a^{n}(t')dt'
\end{equation}
~~~where $m$ and $n$ are constants. This model mimics the general
agegraphic Dark energy models with some differences. The naive
difference backs to the choice of appropriate cut-off of the
model. Also for some reasonable values of $(m,n)$ this model
crosses the phantom line $\omega=-1$. Also the pair $(m,n)$ are
arbitrary and at level of phenomenological the high energy models,
need not to be integers. This model generates ordinary HDE for
$m=n=1$. The new agegraphic corresponds to $(m=0,n=1)$. Finally in
favour of the observational data, the pair $(m,n)=(0,1)$ and
$(m,n)=(4,3)$ are compatible. \\
~~~~ Simple calculation reveals that
\begin{equation}
\dot{L}=-mHL+a^{n-m}(t)
\end{equation}
~~~~~~Now to establish the required correspondence we set
$\rho_{v}=\rho_{G}$. It gives
\begin{equation}
-24H^{3}\dot{G}f_{GG}+Gf_{G}-f(G)=\frac{b^2}{L^2}\left[\frac{2a^{n-m}(t)}{HL}-2m-3\right]
\end{equation}
~~~~ where $f_{G}$, $f_{GG}$ are 1st and 2nd order derivatives of
$f(G)$ with respect to $G$. Now by simple power law representation
$a(t)=a_{0}t^{p}$, where $a_{0}>0$ and $p~>$ are constants we have
the IR cut-off $L=\frac{{a_{0}}^{n-m}}{pn+1}t^{p(n-m)+1}$,
$G=\frac{24(p-1)p^3}{t^4}$. Putting these in (7) and calculating
further we get the equation in terms of $G$ as
\begin{equation}
AG^2f_{GG}-Gf_{G}+f+BG^{\frac{p(n-m)+1}{2}}=0
\end{equation}
where
\begin{equation}
A=\frac{4}{p-1}
\end{equation}
\begin{equation}
B={a_{0}}^{2(m-n)}b^2(1+np)^2\{-3-2m+\frac{2(1+pn)}{p}\}{24p^3(p-1)}^{2p(m-n)-2}
\end{equation}
\\~~~~which is a second order differential equation with solution
\begin{equation}
f(G)=\alpha G^{\frac{1-mp+np}{2}}+c_{1}G^{\frac{1}{A}}+c_{2}G
\end{equation}
where $c_{1}$ and $c_{2}$ are arbitrary integration constants and
\begin{equation}
\alpha=\frac{4B}{(1+mp-np)(-2-A+Amp-Anp)}
\end{equation}
Now we can see that this $f(G)$ increases with $G$ and it forms a
realistic model as in this case $f(G)\rightarrow 0$ as
$G\rightarrow 0$ \cite{51}. In figure 1, we have drawn the $f(G)$
against $G$ for some particular values of $p$, which shows the
ever increasing with positivity nature of $f(G)$ as $G$ increases.

\begin{figure}{h}
\includegraphics[height=2.0in]{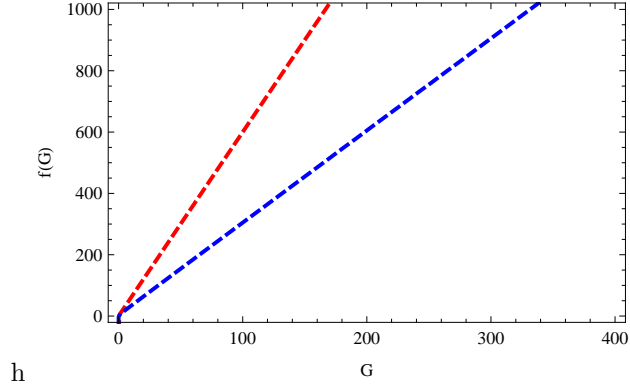}\\
\caption{Evolution of $(m,n)$-type HDE $f(G)$ model versus $G$
with power law scale factor, the Red and Blue lines are associated
with non-zero values of $m,n$ and $p=10$ and $p=2$ respectively.}
\end{figure}

\subsection{\textbf{Reconstruction with entropy-corrected $(m,n)$-type HDE Model}}
Considering the corrected entropy-are relation with derivation of
HDE, we can obtain the energy density of the entropy-corrected
$(m,n)$-type HDE as
\begin{equation}
\rho_{V}=\frac{3c^2}{L^2}+\frac{\xi}{L^4}\log{L^2}+\frac{\zeta}{L^4}
\end{equation}
with $c$, $\xi$, $\zeta$ as dimensionless constants and IR cut-off
as considered in equation (5). Again to reconstruct $f(G)$ with
this dark energy we have considered $\rho_{G}=\rho_{v}$, which
yields
\begin{equation}
\frac{3c^2}{L^2}+\frac{\xi}{L^4}\log{L^2}+\frac{\zeta}{L^4}=-24
H^3 \dot{f_{G}}+Gf_{G}-f
\end{equation}
after considering the power law scale factor $a(t)=a_{0}t^p$, with
$a_{0}>0$ and $p~>1$ constants we can simply discover the equation
(14) as
\begin{equation}
AG^2f_{GG}-Gf_{G}+f+BG^{\frac{pn-pm+1}{2}}+G^{pn-pm+1}(C+D\log{G})=0
\end{equation}
where
\begin{equation}
A=\frac{4}{1-p}
\end{equation}
\begin{equation}
B=3 {a_{0}}^{2(m-n)}c^2(1+pn)^2[24(p-1)p^3]^{\frac{pm-pn+1}{2}}
\end{equation}
\begin{equation}
C={a_{0}^{4(m-n)}}(1+pn)^4 [24 (p-1)p^3]^{pm-pn-1}\left[\zeta +\xi
\log{\{\frac{a_{0}^{2(n-m)}}{(1+pn)^2}{24(p-1)p^3}\}^{\frac{pn-pm+1}{2}}}\right]
\end{equation}
\begin{equation}
D=-\frac{1}{2}{a_{0}^{4(m-n)}}(1+pn)^4[24p^3(p-1)]^{pm-pn-1}\xi^{pn-pm+1}
\end{equation}
which produces $f(G)$ in the form
\begin{equation}
f(G)=\alpha (c_{1}G^{\frac{1}{A}}+c_{2}G)+\beta
G^\frac{1-pm+pn}{2}+G^{1-pm+pn}(\gamma -\delta \log{G})
\end{equation}
where

\begin{figure}
\includegraphics[height=2.0in]{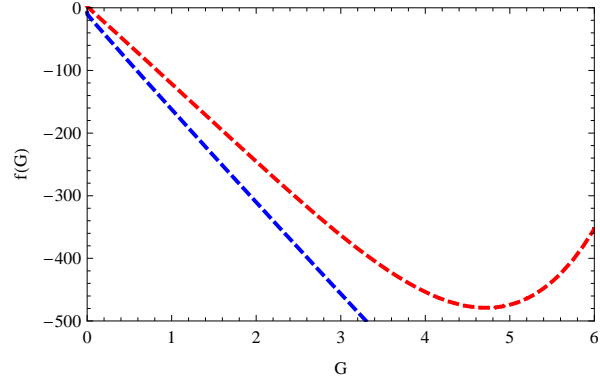}\\
\caption{Evolution of entropy-corrected $(m,n)$-type HDE $f(G)$
model versus $G$ with power law scale factor, the Red and Blue
lines are associated with non-zero values of $m=1.3, n=3, p=2.06$
and $m=0, n=1, p=2$ respectively.}
\end{figure}

\begin{equation}
\tau=(m-n)^2p^2[1+pm-pn(-1+A-Apm+Apn)^2\{2+A(-1+pm-pn)\}]
\end{equation}
\begin{equation}
\alpha=\frac{1}{\tau}(m-n)^{2}p(1+pm-pn)\{1+A(-1+pm-pn)\}^2\{2+A(-1+pm-pn)\}
\end{equation}
\begin{equation}
\beta=\frac{1}{\tau}[-4B(m-n)^2p^2(-1+A-Apm+Apn)^2]
\end{equation}
\begin{equation}
\gamma=\frac{1}{\tau}[(A-1)\{D+C(m-n)\}-A\{2D+C(m-n)\}(m-n)p](-1-pm+pn)(-2+A-Apm+Apn)
\end{equation}
\begin{equation}
\delta=\frac{1}{\tau}D(m-n)p(1+pm-pn)[2+A(-1+pm-pn)\{3+A(-1+pm-pn)\}]
\end{equation}
and $c_{1}$ and $c_{2}$ are arbitrary constants.
\\~~~~We can clearly see by figure 2 that $f(G)$ is a decreasing
function of $G$ and $f(G)\rightarrow 0$ as $G\rightarrow 0$, which
reveals that it is a realistic model.

\section{\textbf{Stability analysis}}
 Squared speed of sound $v_{s}^2=\frac{\dot{p}}{\dot{\rho}}$ is an
 important quantity to test the stability of the background
 evolution. Positive value of $v_{s}^2$ implies classical stability
 of a given perturbation \cite{59,60}. Interacting new HDE is characterized by
 negative $v_{s}^2$ giving classical instability by Sharif et al \cite{59}.
 Myung \cite{591} testified that squared speed for HDE
always  stays at negative level for choosing future event horizon
as IR cut-off, while for those for Chaplygin gas and tachyon it
stays non-negative. Kim et al \cite{60} observed that the perfect
fluid for agegraphic dark energy is classically unstable as
$v_{s}^2$ is always negative and Jawad et al \cite{26} shows that
$f(G)$ model in HDE scenario with power law scale factor is
classically unstable.
 \\~~~~~We here considered the $v_{s}^2$ as equal to
 $\frac{\dot{p}_{eff}}{\dot{\rho}_{eff}}$ and plot this $v_{s}^2$
 verses the cosmic time $t$ for both reconstructions of $f(G)$
 models. For the $(m,n)$-type HDE $f(G)$ model with power law scale
 factor, the model is classically unstable for $m=0, n=1$ (figure 3) and is
 stable for $m=4, n=3$ (figure 4). And for the entropy-corrected $(m,n)$-type
 HDE $f(G)$ model, the model is stable for both $m=0, n=1$ (figure 5) and $m=1,
 n=3$ (figure 6).
 \begin{figure}
\begin{minipage}{16pc}
\includegraphics[width=16pc]{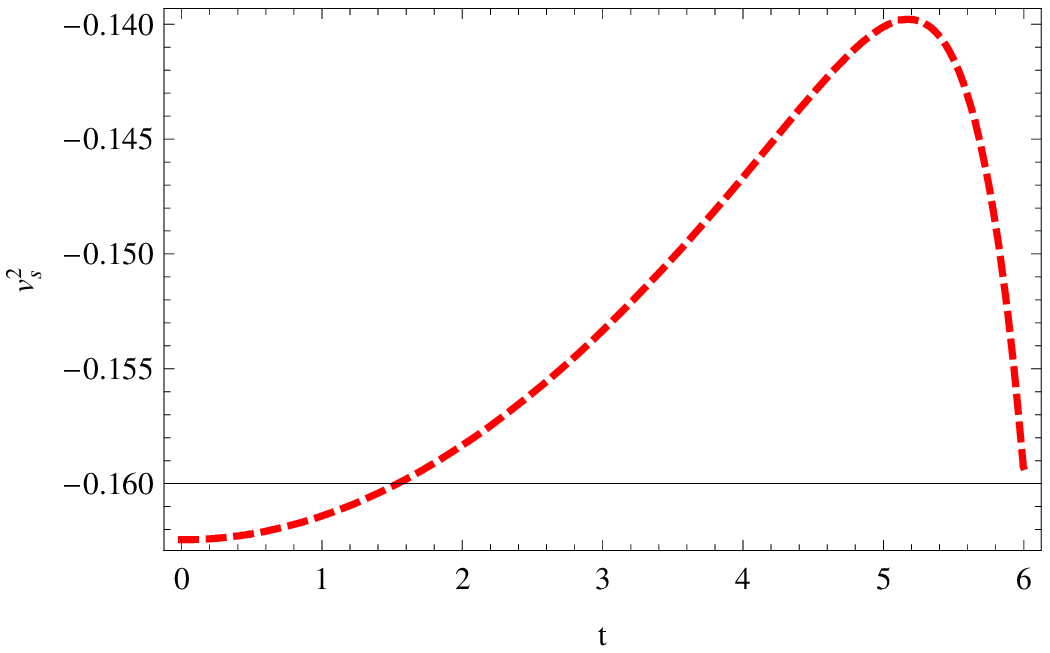}
\caption{\label{label}This figure plots $v_{s}^2$ versus $t$ for
$(m,n)$-type HDE $f(G)$ model and we find that it gives classical
instability for $m=0,n=1$}
\end{minipage}\hspace{3pc}%
\begin{minipage}{16pc}
\includegraphics[width=16pc]{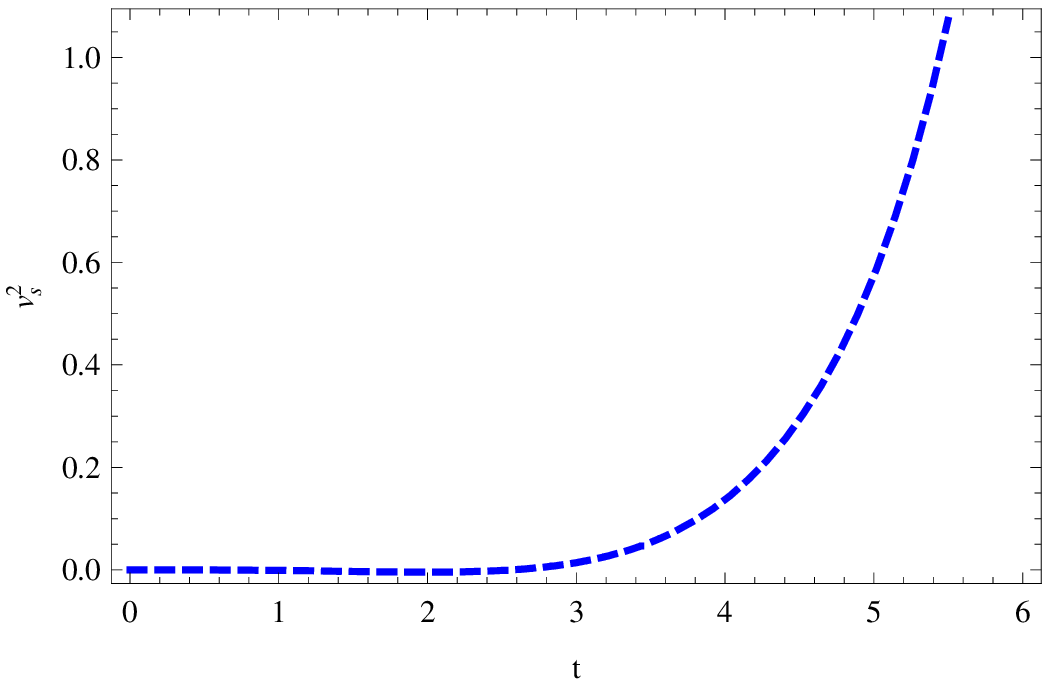}
\caption{\label{label}This figure plots $v_{s}^2$ versus $t$ for
$(m,n)$-type HDE $f(G)$ model and we find that it gives classical
stability for $m=4,n=3$}
\end{minipage}\hspace{3pc}%
\end{figure}
\begin{figure}
\begin{minipage}{16pc}
\includegraphics[width=16pc]{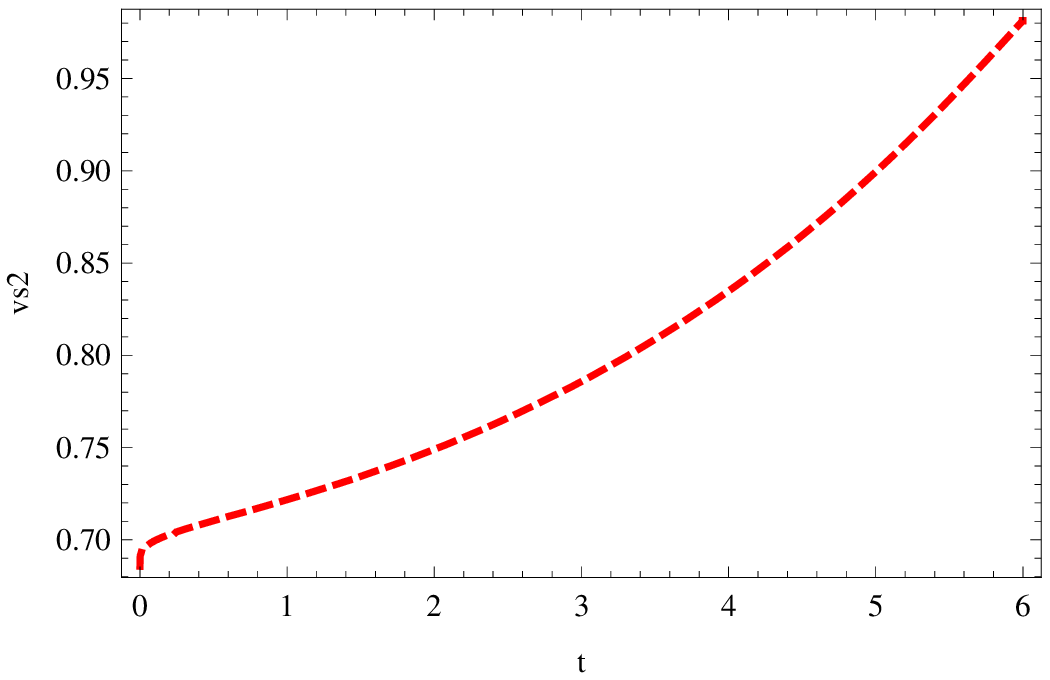}
\caption{\label{label}This figure plots $v_{s}^2$ versus $t$ for
entropy-corrected $(m,n)$-type HDE $f(G)$ model and we find that
it gives classical stability for $m=0,n=1$}
\end{minipage}\hspace{3pc}%
\begin{minipage}{16pc}
\includegraphics[width=16pc]{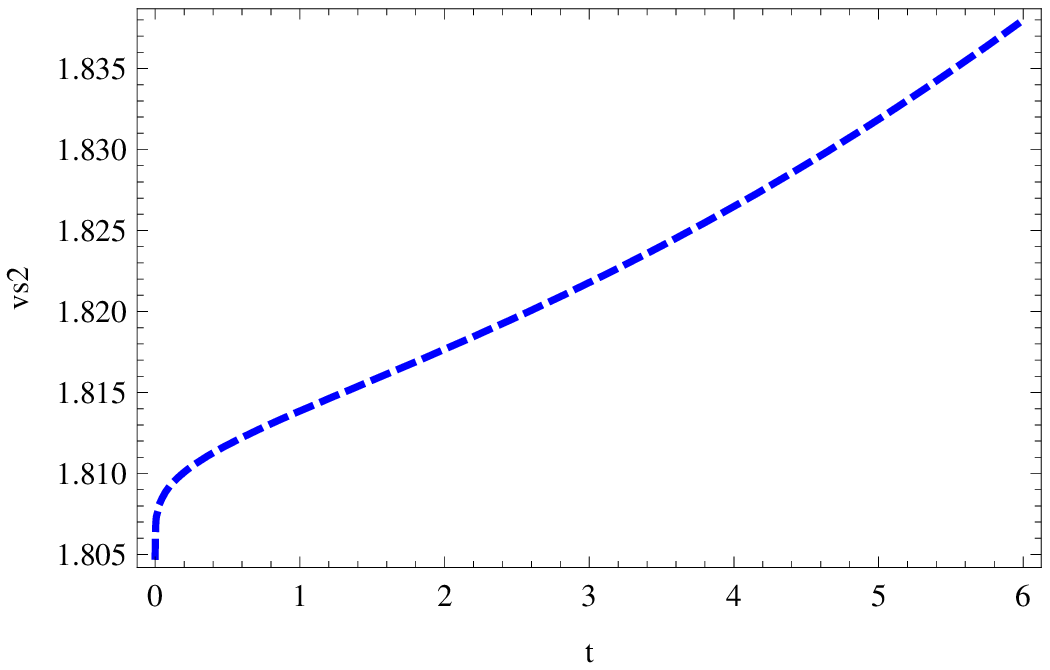}
\caption{\label{label}This figure plots $v_{s}^2$ versus $t$ for
entropy-corrected $(m,n)$-type HDE $f(G)$ model and we find that
it gives classical stability for $m=1,n=3$}
\end{minipage}\hspace{3pc}%
\end{figure}

\section{Concluding remarks}

In this work, we have considered the generalized version of
holographic dark energy (HDE) like $(m,n)$ type HDE model in the
background of $f(G)$ gravity in FRW universe. We have discussed
the correspondence of the well-accepted $f(G)$ gravity theory with
two dark energy models: $(m,n)$-type holographic dark energy
[$(m,n)$-type HDE] and entropy-corrected $(m,n)$-type holographic
dark energy. Here we have considered the power law form of the
scale factor $a(t)=a_{0}t^{p}, p>1$. With this choice, the
explicit forms of $f(G)$ have been found in terms of $G$ for both
models. The model in both cases are found to be realistic i.e.,
$f(G)\rightarrow 0$ as $G\rightarrow 0$. In figures 1 and 2, we
have shown that $f(G)$ decreases with $G$ from positive side for
$(m,n)$-type HDE and from negative side for entropy-corrected
$(m,n)$-type HDE. We have also discussed the classical stability
issues in both models with the condition that $v_{s}^{2}>0$ for
these reconstructed models. For $(m,n)$-type HDE model,
$v_{s}^{2}$ has been drawn in figures 3 and 4 for $(m,n)=(0,1)$
and $(m,n)=(4,3)$. For $(m,n)=(0,1)$, we see that $v_{s}^{2}<0$,
which is unstable but for $(m,n)=(4,3)$, we see that
$v_{s}^{2}>0$, which is classically stable. So we conclude that
$(m,n)$ type HDE is more stable model than ordinary HDE model in
$f(G)$ gravity. Also for entropy-corrected $(m,n)$-type HDE model,
$v_{s}^{2}$ has been drawn in figures 3 and 4 for $(m,n)=(0,1)$
and $(m,n)=(1,3)$. For both the cases, we observe that
$v_{s}^{2}>0$. So entropy-corrected model is classically stable
for both ordinary HDE and $(m,n)$ type HDE models in $f(G)$ gravity.\\

\section{Acknowledgements} The authors sincerely
acknowledge the Visiting program provided by IUCAA, Pune, India in
January 2014 to carry out research in General Relativity and
Cosmology.\\\\

\end{document}